\documentclass[%
 reprint,
 amsmath,amssymb,
 aps,
superscriptaddress, 
]{revtex4-2}

\usepackage{appendix}
\usepackage{graphicx}
\usepackage{dcolumn}
\usepackage{bm}
\usepackage[colorinlistoftodos]{todonotes}
\usepackage{mycomments}
\usepackage[load=physical,load=synchem]{siunitx}

\begin{document}

\title{Fundamental bounds on the precision of iSCAT, COBRI and dark-field microscopy \\ for 3D localization and mass photometry}

\author{Jonathan Dong}%
  \affiliation{Biomedical Imaging Group, École Polytechnique Fédérale de Lausanne, Lausanne 1015, Switzerland}
\author{Dante Maestre}%
  \affiliation{University of Vienna, Faculty of Physics, VCQ, 
A-1090 Vienna, Austria}
  \affiliation{University of Vienna, Max Perutz Laboratories, 
Department of Structural and Computational Biology, A-1030 Vienna, 
Austria}
\author{Clara Conrad-Billroth}%
  \affiliation{University of Vienna, Faculty of Physics, VCQ, 
A-1090 Vienna, Austria}
  \affiliation{Advanced Microscopy Facility, Vienna Biocenter Core Facilities (VBCF), A-1030 Vienna, Austria}
\author{Thomas Juffmann}
\thanks{Corresponding author: thomas.juffmann@univie.ac.at}
  \affiliation{University of Vienna, Faculty of Physics, VCQ, 
A-1090 Vienna, Austria}
  \affiliation{University of Vienna, Max Perutz Laboratories, 
Department of Structural and Computational Biology, A-1030 Vienna, 
Austria}

\date{\today}

\begin{abstract}
Interferometric imaging is an emerging technique for particle tracking and mass photometry. Mass or position are estimated from weak signals, coherently scattered from nanoparticles or single molecules, and interfered with a co-propagating reference. 
In this work, we perform a statistical analysis and derive lower bounds on the measurement precision of the parameters of interest from shot-noise limited images. This is done by computing the classical Cram\'er-Rao bound for localization and mass estimation, using a precise vectorial model of interferometric imaging techniques. We then derive fundamental bounds valid for any imaging system, based on the quantum Cram\'er-Rao formalism. 
This approach enables a rigorous and quantitative comparison of common techniques such as interferometric scattering microscopy (iSCAT), Coherent Brightfield microscopy (COBRI), and dark-field microscopy. In particular, we demonstrate that the light collection geometry in iSCAT greatly increases the axial position sensitivity, and that the Quantum Cram\'er-Rao bound for mass estimation yields a minimum relative estimation error of $\sigma_m/m=1/(2\sqrt{N})$, where $N$ is the number of collected scattered photons.
\end{abstract}

\maketitle



\section{Introduction}
Scattering-based interferometric imaging is a powerful method for the label-free detection and tracking of single biomolecules in solution~\cite{iSCAT_QMI, iSCAT_DOS, iSCAT_DNA,iSCAT_quantMP, iSCAT_tracking, iSCAT_virus, cobri_virus, cobri_trackingofNP}. In these techniques, light scattered by a molecule is interfered with a reference to obtain a detectable signal. Applications include the determination of the mass of biomolecules and their oligomeric states~\cite{iSCAT_QMI, iSCAT_DOS, iSCAT_quantMP, iSCAT_DNA}, as well as high speed tracking of nanoparticles in cellular environments~\cite{iSCAT_tracking, iSCAT_virus, cobri_virus, cobri_trackingofNP}.  

Due to the small scattering cross-section of a single protein (e.g. $10^{-11}\, \micro \mathrm{m}^2$ for bovine serum albumin \cite{iSCAT_DOS}), shot-noise typically limits how precisely mass or position can be estimated. Sensitivity can be improved by increasing the number of collected photons, either via longer observation times or higher incident power. 
However, long observation times limited by temporal dynamics of the sample and setup instabilities, and incident power is limited by detector well-depth and adverse photo-induced heating. 
It is therefore crucial to extract as much information as possible from every scattered photon.

The Cramér-Rao bound (CRB) quantifies the maximal achievable precision regarding the estimation of parameters from noisy measurements~\cite{Trees2013DetectionI, Barrett2013FoundationsScience}. CRBs have been used to derive the achievable precision regarding localization~\cite{Chao:16, Shechtman2014OptimalImaging,balzarotti_nanometer_2017} and lifetime ~\cite{bouchet2019cramer} estimation in fluorescence microscopy, as well as phase and amplitude estimation in coherent microscopy techniques~\cite{ PhysRevApplied.15.024047, Koppell:21}. Furthermore, quantum estimation theory~\cite{tsang2016quantum, backlund2018fundamental} provides fundamental bounds on precision, describing the amount of information which can be extracted from the quantum state of light itself. One can thus compare the classical CRB, achieved with a given measurement system, with the Quantum CRB (QCRB), bounding the achievable precision for \textit{any} measurement system. 

Here, we apply this formalism to scattering-based microscopy techniques to quantify the estimation precision regarding particle mass and position. We specifically compare interferometric scattering microscopy (iSCAT) \cite{iSCAT_rev_kukura, iSCAT_rev_sandoghdar}, Coherent Brightfield Microscopy (COBRI)~\cite{HSIEH201869, cobri_rev}, and Dark-Field microscopy (DF)~\cite{doi:10.1021/ph500138u}. In Section~\ref{sec: theoretical background}, we first adapt a vectorial model of image formation~\cite{Torok:95, haeberle2003focusing, Aguet:09}, already applied for iSCAT~\cite{GholamiMahmoodabadi:20}, to COBRI and DF. We then introduce a numerical model for the calculation of CRBs, and finally derive analytic expressions for the corresponding QCRBs.

We apply this theoretical framework to interferometric imaging in Section~\ref{sec: results}. First, we discuss nanoparticle localization, and show that all three techniques yield similar CRBs for transverse $x, y$ localization, while iSCAT can be more than $5 \times$ more precise for $z$ localization. 
Second, we calculate mass estimation precision, and show that DF is slightly more photon-efficient than interferometric techniques. 
Third, we discuss how these bounds are affected by attenuation of the reference light \cite{cobri_trackingofNP, iSCAT_attenuator_kukura, iSCAT_amplified}.
\section{Theory - background and results}
\label{sec: theoretical background}
In this section, we first introduce the vectorial imaging model to precisely calculate the scattered fields. We then describe the concept of classical Fisher Information (FI) and CRBs for shot-noise limited measurements, followed by their quantum counterparts. 

\subsection{Image formation}

\begin{figure}[t]
    \centering
    \includegraphics[width=\columnwidth]{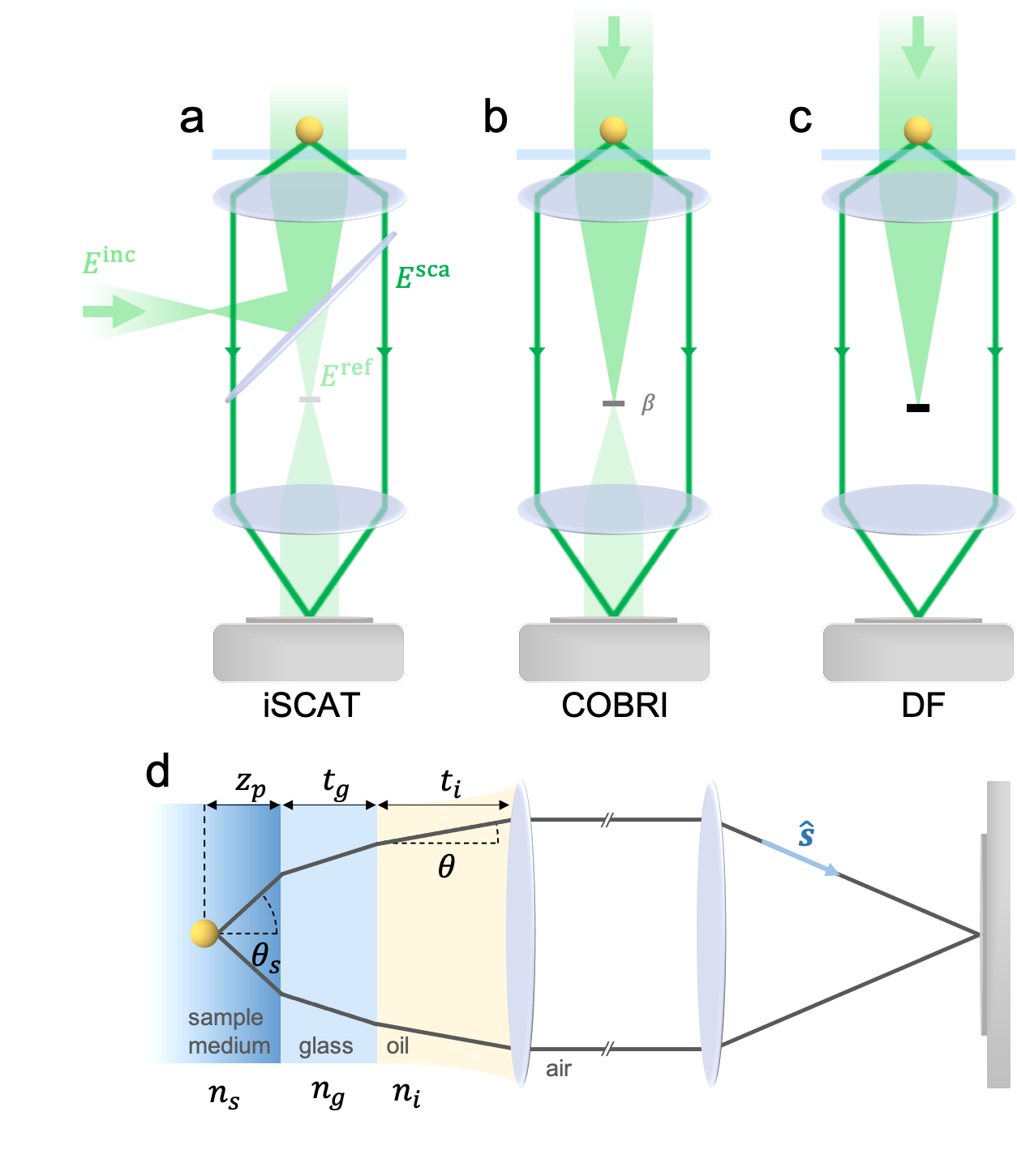}
    \caption[Description]{Sketches of the experimental configuration in (a) iSCAT, (b) COBRI and (c) DF microscopy. The light scattered by the particle on the coverslip is imaged onto a camera where it interferes with a reference. In iSCAT, the reference corresponds to the light reflected at the coverslip-sample interface. In COBRI the transmitted light is used as a reference, which can be selectively attenuated in the back-focal plane of the objective lens by a factor $\beta$. In DF the reference light is blocked completely, and only the scattered light is detected. In (d) the typical three-layer structure is depicted, showing the nanoparticle within a sample medium, a glass coverslip, and the immersion oil. Aberrations are introduced when the experiment deviates from the design conditions of the microscope objective.} 
    \label{fig:setup}
\end{figure}

In this study we focus on three coherent microscopy techniques, as illustrated in Fig.~\ref{fig:setup}. 
In all considered geometries, incident light of wavelength $\lambda$ induces dipole scattering. In iSCAT microscopy, the backwards-scattered light is interfered with the portion of the incident light that is reflected at the water-glass interface. In COBRI, the forward scattered light is interfered with the transmitted incident light. In order to increase contrast, the reference light is often selectively attenuated by a factor $\beta$ using a mask in the Fourier plane \cite{cobri_trackingofNP}. In DF, the scattered light is detected without interfering it with a reference field. 
To compare these different imaging schemes, a detailed model of image formation is required. 
Here, we apply a vectorial model~\cite{Torok:95, https://doi.org/10.1046/j.1365-2818.1999.00462.x, haeberle2003focusing, Aguet:09} that goes beyond the paraxial approximation in order to account for light collection with high numerical aperture (NA). The model has recently been applied to iSCAT~\cite{GholamiMahmoodabadi:20}, and demonstrated the importance of aberrations introduced by particle defocus. In the following, we lay out the most important steps of our model, which also applied to DF and COBRI. Details can be found in the appendix.

The detected intensity can be calculated as the interference of the scattered field $\boldsymbol E^{\rm{sca}}$ with a reference field $\boldsymbol E^{\rm{ref}}$ at the detector plane:
\begin{equation}
I^{\rm{det}} = 
    \big \vert \boldsymbol E^{\rm{ref}} \big \vert^2 + 
    \big \vert \boldsymbol E^{\rm{sca}} \big \vert^2 + 
    2 \big \vert \boldsymbol E^{\rm{ref}} \big \vert 
        \big \vert \boldsymbol E^{\rm{sca}} \big \vert 
        \cos{\phi^{\rm{sca}}}
\label{eq: expected_intensity}
\end{equation} 
where $\phi^{\rm{sca}}$ corresponds to the phase difference between the two fields. Note that in our notation we omit the spatial dependence and the incoherent sum on the two components of the electric field for brevity. 

The calculation of the amplitude of the reference light is straightforward: assuming a linearly-polarized incident field $\boldsymbol{E}^{\mathrm{inc}} = E^{\mathrm{inc}} \hat{\boldsymbol{e}}_x$, we get $\boldsymbol{E}^{\rm{ref}} = r_\parallel \boldsymbol{E}^{\mathrm{inc}}$ for iSCAT, and $\boldsymbol{E}^{\rm{ref}} = \beta t_\parallel \boldsymbol{E}^{\mathrm{inc}}$ for COBRI. Here, $r_\parallel$ and $t_\parallel$ are the Fresnel coefficients for the reflection and transmission of $p$-polarized light at the coverslip-water interface ($r_\bot$ and $t_\bot$ will be used for $s$-polarized light later), and $\beta$ represents an optional attenuation factor. For DF, $\boldsymbol{E}^{\rm{ref}}=0$.

The calculation of the scattered field is more involved: we start with the scattering amplitude $\boldsymbol{E}^1 \equiv E^1 e^{i \psi_0} \hat{\boldsymbol{e}}_x$, which is proportional to the complex polarizability $\alpha$ of the particle and the incident field: $\boldsymbol{E}^1 \propto \alpha \boldsymbol{E}^{\mathrm{inc}}$. Next, we need to include aberrations with a precise propagation model. We assume that a microscope objective provides aberration-free images under very precise design settings: the imaged plane must be located at the surface of a coverslip of thickness $t_g^*$ and refractive index $n_g^*$, after an immersion oil layer of thickness $t_i^*$ and refractive index $n_i^*$. Deviations from those settings will cause aberrations. As shown in Fig.~\ref{fig:setup}d, we now consider a particle at a distance $z_p$ from the coverslip, and an immersion layer of thickness $t_i$, which is the typical experimental configuration as described in~\cite{Gibson:92}. The indices of refraction of the sample medium, the immersion layer, and the cover glass are denoted by $n_s$, $n_i$, and $n_g$ respectively. We assume for simplicity that $t_g = t_g^*$, $n_g = n_i = n_i^* = n_g^*$. $\theta_s$ and $\theta$ are the angles of the optical rays with respect to the optical axis in the sample medium and immersion medium, respectively. We also define the maximal angular aperture of the objective $\alpha_a = \sin^{-1}(\mathrm{NA}/n_i)$, the azimuthal angle $\phi$ of a unit vector $\hat{\boldsymbol{s}}$, and the detector position $\boldsymbol{r}$, with the origin at the central focus position in the design setting.

With these settings, the scattered field in the detector plane is given by a Richards-Wolf integral \cite{Torok:95, Aguet:09, richards1959electromagnetic}:
\begin{equation}
    \boldsymbol{E}^{\rm{sca}} = - \frac{i}{\lambda} E^1 e^{i \psi} 
    \int_0^{\alpha_{a}} \int_0^{2 \pi} \boldsymbol{A}
    e^{i k \Lambda}
    e^{i k \hat{\boldsymbol{s}} \cdot \boldsymbol{r}}
    \sin \theta \sqrt{\cos \theta} d\theta d\phi
    \label{eq: rw integral}
\end{equation}
with $\boldsymbol{A}$ a vector defined as:
\begin{equation}
    \boldsymbol{A} =
    \begin{pmatrix}
        t_\parallel \cos \theta_s \cos^2 \phi + t_\bot \sin^2 \phi \\
        (t_\parallel \cos \theta_s - t_\bot) \cos \phi \sin \phi
    \end{pmatrix}
\end{equation}
and the aberration term $\Lambda$ depending on the geometry:
\begin{equation}
    \Lambda = z_p n_s (\cos \theta_s+\xi) + n_i (t_i-t_i^*) (\cos \theta - 1)
    \label{eq: aberration term iSCAT}
\end{equation}
where $\xi=+1$ for iSCAT, and $\xi=-1$ for COBRI. This difference is a result of an additional optical path length in iSCAT since the reference light and the scattered light do not originate from the same plane as compared to COBRI. A more compact formalism used to accelerate the numerical study is presented in the Appendix.

\subsection{Cram\'er-Rao bounds introduction}

From measuring the spatial distribution of intensities $I^{\rm{det}}(\boldsymbol{r})$ one can estimate unknown parameters $\gamma$, like the position $x, y, z$, or mass $m$ of a particle. Any noise in the measurement will inevitably lead to stochastic estimations. The information about $\gamma$, which is contained in the detected, noisy intensities, can be quantified using the concept of Fisher Information (FI)~\cite{Trees2013DetectionI}. We build the FI matrix by computing the FI for each pair of parameters $\gamma_i$ and $\gamma_j$, for the case where shot-noise is the dominant source of noise~\cite{PhysRevApplied.15.024047} by:
\begin{equation}
     [\mathcal{J}(\gamma)]_{ij} 
     = 
     \int d\boldsymbol{r}
     \frac{1}{I^{\rm{det}}(\boldsymbol{r})}
     \left( \frac{\partial I^{\rm{det}}(\boldsymbol{r})}{\partial \gamma_i} \right)
     \left( \frac{\partial I^{\rm{det}}(\boldsymbol{r})}{\partial \gamma_j} \right) \; .
     \label{eq: FI}
\end{equation}

The variance of any unbiased estimator $\hat{\gamma}(I^{\rm{det}})$ of the parameters $\gamma$ must satisfy the Cram\'{e}r-Rao inequality~\cite{Trees2013DetectionI} given by: 
\begin{equation}
    \operatorname{Var}(\hat{\gamma}_j) \geq [ \mathcal{J}^{-1} (\gamma) ]_{jj} \geq \frac{1}{[\mathcal{J} (\gamma)]_{jj}} \; .
    \label{eq: crb ineq}
\end{equation}
The first inequality gives the CRB for estimation precision of $\gamma_j$, when all other parameters are unknown and the estimation is potentially affected by crosstalk. The second inequality yields the CRB assuming perfect knowledge of all other parameters. Finally, the corresponding lower bound on standard deviation is:
\begin{equation}
    \sigma_{\mathrm{CRB},\,\gamma_j} = \frac{1}{\sqrt{[\mathcal{J} (\gamma)]_{jj}}}
\end{equation}

\subsection{Quantum Cram\'er-Rao bounds}

The Quantum Fisher Information (QFI) and the associated Quantum Cram\'er-Rao Bounds (QCRB) provide fundamental estimation bounds which are valid for \textit{any} measurement system~\cite{helstrom1976quantum}. The QFI quantifies the information contained in the quantum state of light itself. 
Recently this approach was applied to localization estimation precision in fluorescence microscopy~\cite{Shechtman2014OptimalImaging}, where the parameters of interest were given by the position of the particle $(\gamma_1, \gamma_2, \gamma_3)=(x_p,y_p,z_p)$, and the state of the fluorescence light was defined as a superposition of single photon states described by their coordinates in Fourier space. In interferometric scattering microscopy, the mass of the particle constitutes a further parameter of interest $\gamma_4=m$, and the light scattered by a particle can be described as a superposition of coherent states parametrized by $(\theta, \phi)$ with amplitudes $\epsilon(\theta, \phi)$ described in Eq.~\eqref{eq: rw integral}. Note that Eq.~\eqref{eq: expected_intensity} assumes coherence between the reference and the scattered light, which constitutes a fundamental difference to the treatment of incoherent fluorescent light~\cite{backlund2018fundamental}. These notions allow calculating the QFI of the scattered fields collected by the objective, and the QCRBs yielding a bound on estimation precision irrespective of the measurement scheme that follows. 
\begin{figure*}[t]
    \centering
  \includegraphics[width=18cm]{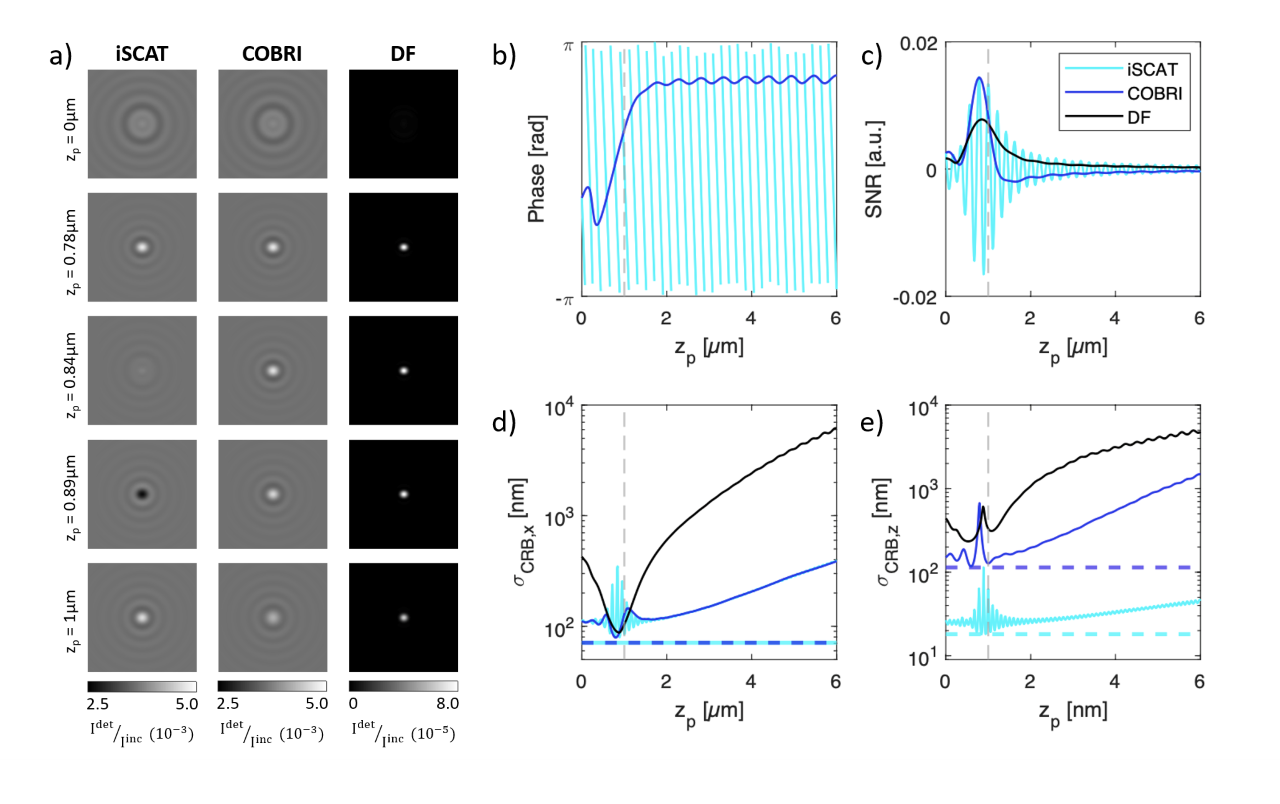}
    \caption[Description2]{(a) Simulated detected intensity for iSCAT, COBRI, and DF at different axial particle positions $z_p$. The field of view is $4 \times 4$~{\textmu}m$^2$. (b) On-axis phase shift $\phi^{\rm{sca}}(z_p)$ for iSCAT (cyan) and COBRI (blue). (c) On-axis signal to noise $\text{SNR}(z_p)$ for iSCAT, COBRI, and DF (black). (d) Transverse and (e) axial CRBs for the different schemes as a function of $z_p$, normalized for one scattered photon detected. The dashed horizontal lines in panels (d) and (e) indicate the respective quantum CRBs. For all panels the focus plane $z_f$ was set to $1$~{\textmu}m and is indicated by the horizontal dashed line.
    }
    \label{fig: localization_precision}
\end{figure*}
Following the derivation in~\cite{bouchet2021maximum}, we can write the QFI as:
\begin{equation}
    \label{eq: qcrb general}
    \mathcal{K}_{jj} = 
        4 \int_0^{\alpha_{a}} d\theta \int_0^{2\pi} d\phi
        \left|\partial_j \epsilon(\theta, \phi)
        \right|^2
\end{equation}
The QFIs for localization, normalized for one collected scattered photon, are given by: 
\begin{align}
    \label{eq: qcrb x}
    \mathcal{K}_{xx} &= 
        \frac{1}{\mathcal A} \int_0^{\alpha_{a}} d\theta \int_0^{2\pi} d\phi
        \left|k n_i \cos \phi \sin^2 \theta 
        \sqrt{\cos \theta} \boldsymbol{A}
        \right|^2 \\
    \mathcal{K}_{yy} &= 
        \frac{1}{\mathcal A} \int_0^{\alpha_{a}} d\theta \int_0^{2\pi} d\phi
        \left|k n_i \sin \phi \sin^2 \theta
        \sqrt{\cos \theta} \boldsymbol{A}
        \right|^2 \\
          \label{eq: qcrb z}
    \mathcal{K}_{zz} &= 
        \frac{1}{\mathcal A} \int_0^{\alpha_{a}} d\theta \int_0^{2\pi} d\phi
        \left|k n_s (\cos \theta_s + \xi)
        \sin \theta \sqrt{\cos \theta} \boldsymbol{A}
        \right|^2
\end{align}
where $\cos\theta_s = \sqrt{1-n_s^2\sin^2\theta/n_i^2}$. Note that Eq.~\eqref{eq: qcrb z} depends on $\xi$, i.e. on the optical path length difference of reference and scattered light. $\mathcal A$ is a normalization factor given by:
\begin{equation}
    \mathcal A =
        \int_0^{\alpha_{a}} d\theta \int_0^{2\pi} d\phi
        \left|\sin \theta 
        \sqrt{\cos \theta} \boldsymbol{A}
        \right|^2
\end{equation}

The QFI for mass estimation is:
\begin{equation}
    \mathcal{K}_{mm} = \frac{4}{m^2}
\end{equation}

The associated QCRBs are:
\begin{align}
    \operatorname{Var}(\hat{\gamma}_j) \geq \sigma_{\mathrm{QCRB},\,\gamma_j}^2 = \frac{1}{\mathcal{K}_{jj}}
    \label{eq: qcrb ineq}
\end{align}
Written for the single parameter estimation, these fundamental bounds also bound the maximal achievable precision for multi-parameter estimation. In particular, this leads to the following bound on relative mass estimation:
\begin{equation}
    \frac{\sigma_{\mathrm{QCRB},\,m}}{m} = \frac{1}{2}
\end{equation}

These bounds have been derived for one scattered photon collected by the optical system. To get the bounds for $N$ photons the variances have to be scaled by $1/N$, and standard deviations by $1/\sqrt{N}$.

\section{Numerical results}
\label{sec: results}

\subsection{3D localization precision}

Interferometric imaging allows precise localization of single nanoparticles. Since the number of scattered photons is only limited by the intensity of the incoming field, precise tracking can be achieved on fast timescales and over an extended period~\cite{iSCAT_tracking}. This is a unique advantage compared to fluorescence microscopy, where finite lifetimes and bleaching limit tracking speed, precision and observation time. 

Applying the equations introduced in the previous sections, we can study the performance of interferometric scattering techniques regarding localization precision. Fig.~\ref{fig: localization_precision} presents the results obtained for a $30\,\mathrm{nm}$-diameter gold nanoparticle. The microscope is focused to $1\,\mathrm{\mu m}$ above the cover glass. All other relevant parameters for the simulation can be found in the appendix. 
Fig.~\ref{fig: localization_precision}(a) shows the simulated intensities (normalized to the incident intensity $I^{\mathrm{inc}}$) for iSCAT, COBRI (for $\beta t_\parallel=r_\parallel$), and DF. Intensities are calculated for 5 different axial particle positions $z_p $ and we observe that the signal modulations are larger for iSCAT and COBRI than for DF, due to the interference term in Eq.~\eqref{eq: expected_intensity}. A slight azimuthal asymmetry is observed due to polarization effects. Note that the $z$-positions of the particle were not chosen at equidistant steps in order to highlight the fast oscillations in the modulation of the iSCAT signal, as reported in~\cite{GholamiMahmoodabadi:20}. 

These fast oscillations in iSCAT are due to fast changes in the phase between the reference and the scattered light (plotted in Fig.~\ref{fig: localization_precision}(b)), arising from changes in the relative optical path length that scale linearly with $z_p$. This is not observed for COBRI, where both fields originate from the same plane. For both techniques, there is an additional geometric phase shift, i.e. the Gouy phase shift. While a Gouy phase shift of $\pi$ would be expected for a spherical wave passing through a focus~\cite{Feng:01}, this can be drastically different for highly aberrated beams~\cite{Pang2013, Pang2014}, which needs to be considered in the analysis of 3D tracking data.

Fig.~\ref{fig: localization_precision}(c) plots the on-axis (= center pixel) signal to noise ratio (SNR) for the three techniques as a function of $z_p$. Assuming shot-noise limited measurements, the SNR is proportional to:
\begin{equation}
\text{SNR} \propto \frac{\left \vert \boldsymbol E^{\rm{sca}} \right \vert^2 + 2 \left \vert \boldsymbol E^{\rm{ref}} \right \vert \left \vert \boldsymbol E^{\rm{sca}} \right \vert \cos{\phi^{\rm{sca}}}}{\sqrt{ \left \vert \boldsymbol E^{\rm{ref}} \right \vert^2 +\left \vert \boldsymbol E^{\rm{sca}} \right \vert^2 + 2 \left \vert \boldsymbol E^{\rm{ref}} \right \vert \left \vert \boldsymbol E^{\rm{sca}} \right \vert \cos{\phi^{\rm{sca}}}}}
\end{equation}
where we set $\left \vert E^{\rm{ref}} \right \vert =0$ for DF. The proportionality constant depends on the collection efficiency and is the same for all methods discussed here.
We see that the changes in phase lead to fast oscillations of the on-axis iSCAT SNR. For all three techniques, the maximum SNR values are not observed in focus, but closer to the cover glass. 
SNR values in iSCAT and COBRI are twice as high as in DF, which agrees with the analytic expression, where one obtains $\text{SNR}\propto\left \vert  E^{\rm{sca}} \right \vert$ for DF and $\text{SNR}\propto 2 \left \vert E^{\rm{sca}} \right \vert \cos{\phi}$ for iSCAT and COBRI, as long as $\left \vert E^{\rm{ref}} \right \vert \gg \left \vert E^{\rm{sca}} \right \vert$. 

From this analysis one might conclude that iSCAT and COBRI enable a localization precision twice as high as DF. This notion however does not take the full PSF of the microscope into account. We consider the task of estimating the transverse and axial positions of the particle from the full PSF information, and compute the associated CRBs of Eq.~\eqref{eq: crb ineq} using finite differences.
The results are shown in Fig.~\ref{fig: localization_precision}(d,e), and are given per scattered photon that is collected. For a finite number $N$ of scattered photons collected, the CRBs and QCRBs scale as $1/\sqrt{N}$. Note that we plot standard deviations, i.e. the square root of the CRB and QCRB of Eqs.~\eqref{eq: crb ineq} and~\eqref{eq: qcrb ineq}.

We observe that the minimal CRB for localization precision in the transverse $x$-direction is similar for all three techniques (in $y$-direction it would differ slightly, due to the asymmetry introduced by polarization). The CRB for interferometric measurements fluctuates, depending on the phase at the center of the PSF. These fluctuations are more rapid for iSCAT than COBRI. The axial dependence of the transverse localization precision shows pronounced differences, with iSCAT and COBRI offering higher precision than DF over a larger defocus range. This is a consequence of the more pronounced off-axis features of the respective PSFs, and is especially important for three dimensional tracking applications. 

\begin{table}[t!]
    \centering
    {\setlength{\extrarowheight}{2pt}%
    \begin{tabular}{ m{2cm}|m{2cm}|m{2cm} }
    & \textbf{iSCAT} & \textbf{COBRI} 
    \\[2pt]
    \hline 
    $\sigma_{\mathrm{QCRB},\,x}$ & 71 nm & 71 nm \\
    $\sigma_{\mathrm{QCRB},\,y}$ & 57 nm & 57 nm \\
    $\sigma_{\mathrm{QCRB},\,z}$ & \textbf{18 nm} & 114 nm
    \end{tabular}}
    \caption{Quantum Cram\'er-Rao Bounds for 3D localization in iSCAT and COBRI, normalized for one scattered photon collected.}
    \label{table: qcrb}
\end{table}

Fig.~\ref{fig: localization_precision}(e) presents the axial localization precision of the three techniques. COBRI and iSCAT yield higher axial precision compared to DF thanks to the phase sensitivity of the interferometric imaging scheme. Since this phase varies much faster for iSCAT than for COBRI (Fig.~\ref{fig: localization_precision}(b)), the axial localization precision is greatly improved throughout the entire $z_p$ range. Geometrically this has been explained in the discussion of Fig.~\ref{fig: localization_precision}b, mathematically it is represented by the $\xi = \pm 1$ term in Eq.~\eqref{eq: aberration term iSCAT} affecting the $z$ derivative as seen in Eq.~\eqref{eq: qcrb z}. 
This significant difference in sensitivity is a particularly promising feature of iSCAT, since axial localization is a notoriously challenging problem for which double-objective collection~\cite{backlund2018fundamental} or PSF engineering~\cite{Shechtman2014OptimalImaging} have been proposed. 

The corresponding QCRBs are depicted as horizontal dashed lines in Fig.~\ref{fig: localization_precision}(d,e) and their values are given in Table~\ref{table: qcrb}. They are derived from the quantum state of light itself, for the two different geometries of iSCAT and COBRI. We see that for $x$ localization, the QCRBs in both cases are equal and that the optimal CRB for all three techniques saturate this fundamental bound. On the other hand, for $z$ localization, the QCRB for iSCAT is significantly lower than for COBRI. Both interferometric techniques saturate this bound when the particle is close to the focus position. On the other hand, the axial sensitivity of DF does not reach the QCRB.
Thus, the addition of a copropagating reference provides an effective method to saturate the QCRB, comparable to other interferometric strategies~\cite{backlund2018fundamental} or PSF engineering~\cite{nehme2020deepstorm3d}.

\subsection{Mass estimation}

Mass photometry is one of the most widespread applications of iSCAT~\cite{iSCAT_QMI, iSCAT_DNA, iSCAT_DOS}. It relies on the fact that the polarizability $\alpha$ is proportional to the mass of the nanoparticle. 
The task of estimating the mass of the particle becomes a task of estimating the amplitude of the scattered light. 

\begin{figure}[t]
    \centering
    \includegraphics[width=0.8\columnwidth]{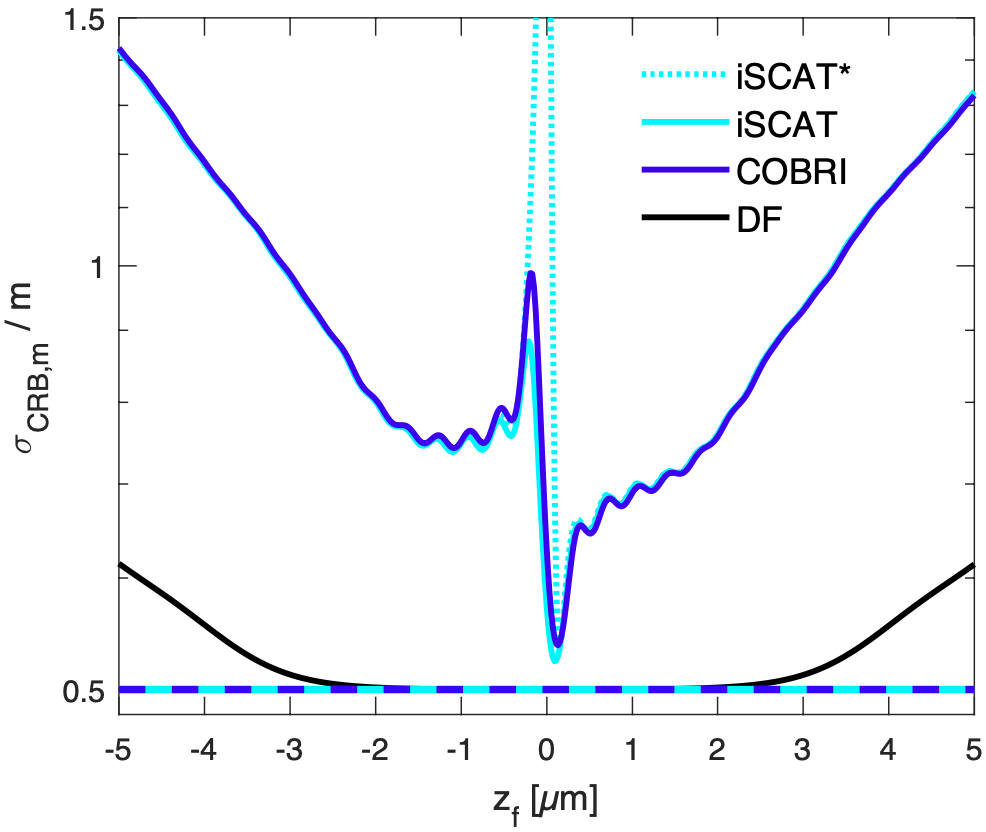}
    \caption[Description]{Cram\'er-Rao bounds for mass estimation, as a function of $z_f$ for iSCAT (cyan), COBRI (blue), and Darkfield (black). The dotted line gives the CRB for the joint estimation of mass and $z_p$ position in iSCAT. The CRBs are normalized by the mass to present relative precision. Quantum Cram\'er-Rao bounds correspond to the horizontal dashed lines, they do not depend on the imaging geometry. The particle is sitting on the coverslip at $z_p = 5$~nm.
    }
    \label{fig: mass_theory}
\end{figure}

The highest precision in mass estimation using interferometric techniques is typically achieved when the particle binds to the cover glass during the observation. Frames recorded before the binding event can then serve as an accurate estimation of the background. This enables extensive averaging and leads to measurements unaffected by nonspecific scattering background. Assuming perfect background subtraction, Fig.~\ref{fig: mass_theory} shows the CRB for mass estimation as a function of the focus position of the microscope. While COBRI and iSCAT yield a similar precision, DF slightly surpasses these two techniques. All three techniques have the same QCRB, which only depends on the scattered amplitude. This means that DF is, in principle, the most efficient strategy for mass estimation among them. Experimentally, however, mass estimation using DF is difficult. This is due to the much smaller signal (see Fig.~\ref{fig: localization_precision} a)), which makes DF more prone to other noise sources (e.g. camera read noise), as well as to systematic errors due to spurious reflections and scattered fields from elements along the optical system~\cite{iSCAT_rev_sandoghdar}. 

The dotted line in Fig.~\ref{fig: mass_theory} shows the mass estimation precision obtained for iSCAT, when both $m$ and $z_p$ are unknown. In this case the off-diagonal terms of the FI matrix contribute significantly to the CRB, leading to lower estimation precision. In practice this means that slight alterations in $z_p$ or $m$ can lead to similar changes in the PSF, and are therefore difficult to discern. This degeneracy is lifted if the particle is moved out of focus, where the shape of the PSF allows for an efficient estimation of both parameters. 

The CRB on mass estimation increases as we move away from the focus position. This is due to the finite field-of-view of our simulations ($4 \times 4$~{\textmu}m), since energy is lost outside the field-of-view for large $z_f$ values. While a larger field of view could be considered, this can become challenging experimentally, especially in the presence of other scatterers, additional noise sources, or an inhomogeneous reference wave. 

\subsection{Attenuation of the reference beam}

\begin{figure}[t]
    \centering
    \includegraphics[width=\columnwidth]{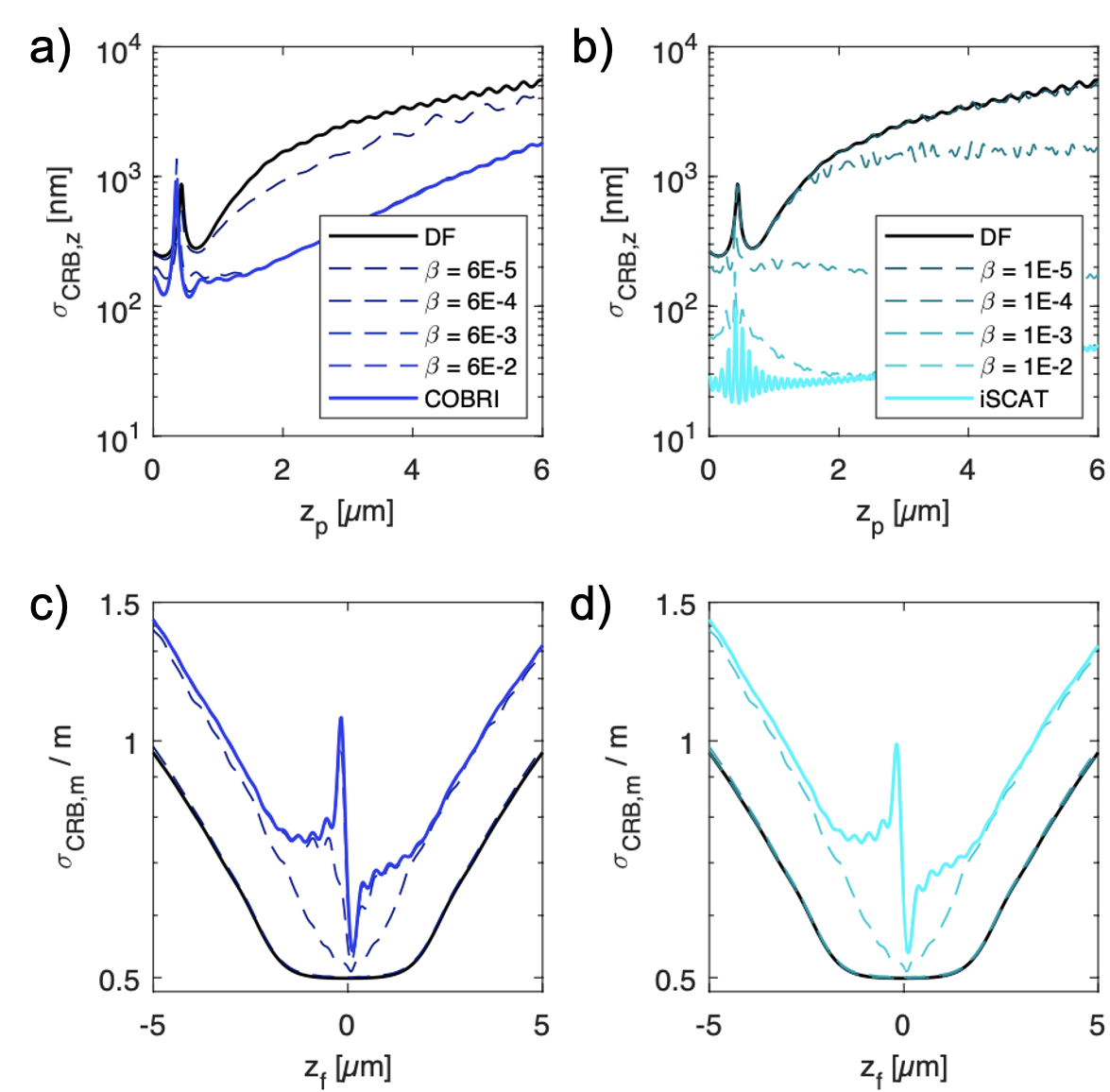}
    \caption[Description]{Cram\'er-Rao bounds for different attenuation values $\beta$. (a) and (b) depict the CRB for the estimation precision of the axial particle position $z_p$ for COBRI and iSCAT at varying attenuation (dashed lines) as a function of $z_p$. The same parameters as in Fig. 2 were used. In (c) and (d) the change in CRB for mass estimation as a function of $z_f$ are illustrated for COBRI and iSCAT for increasing attenuation. The same parameters as in Fig. 3 were used.}
    \label{fig: attenuation}
\end{figure}

Several groups specifically attenuate the reference light in order to increase the contrast in iSCAT and COBRI \cite{cobri_trackingofNP, iSCAT_attenuator_kukura, iSCAT_amplified}, which decreases the requirements on detector well-depth and frame rate. 
To investigate the impact of attenuation  on localization and mass estimation precision we plot the respective CRBs for different attenuation values in Fig.~\ref{fig: attenuation}. We assume that the reference light can be attenuated without affecting the scattered light. Experimentally this can be realized using a mask in the back focal plane of the objective as depicted in Fig.~\ref{fig:setup}.

We observe that an increasing attenuation leads to a monotonous transition between an interferometric setup into a DF setup, with all the consequences discussed in the previous sections. Both for iSCAT and COBRI, we see that the achieved CRBs do not change as long as $\left \vert \boldsymbol E^{\rm{ref}} \right \vert \gg \left \vert \boldsymbol E^{\rm{sca}} \right \vert$, and that we retrieve the DF characteristics for $\left \vert \boldsymbol E^{\rm{ref}} \right \vert \ll \left \vert \boldsymbol E^{\rm{sca}} \right \vert$. We deduce that attenuation in iSCAT should be avoided if axial localization is of concern. For transverse localization and mass estimation, other experimental constraints might outweigh the gains and losses in QFI offered by the introduction of attenuation.



\section{Discussion and Outlook}

In this work we derived the estimation precision regarding particle mass and localization that can be achieved in scattering-based imaging schemes. Specifically, we compared interferometric scattering microscopy (iSCAT), Coherent Brightfield Microscopy (COBRI), and Dark-Field microscopy (DF) and calculated the respective classical and quantum Cram\'er-Rao Bounds ((Q)CRB) for shot-noise limited measurements. 


Most notably, we find that iSCAT yields a significantly better axial localization precision than COBRI and DF. This is due to the information related to the relative phase accumulated between reference and scattered fields, which translates into oscillations in iSCAT signal as a function of particle position. For this reason, iSCAT also performs significantly better than single-objective fluorescence microscopy~\cite{backlund2018fundamental}.
This information advantage also translates to the QCRB, in principle enabling an estimation error roughly 5 times smaller than those in COBRI and DF. 

Regarding mass photometry, we find that all three techniques offer the same QCRB, since mass estimation is tightly related to estimating the amplitude of the scattered light and not its phase. While DF yields a lower CRB than iSCAT and COBRI, this advantage is hard to leverage experimentally, since the small signals in DF require low-noise detection, and excellent suppression of spurious signals. Note that our CRB-results for DF also apply to other DF geometries like coherent oblique angle light sheet microscopy~\cite{Bishop:20}. We also show that there is a divergence in mass photometry precision in iSCAT, if the particle is close to the focus plane, and if neither the mass, nor $z_p$ are known. Experimentally the divergence can be avoided by a slight defocus of the sample. From the QCRB, a fundamental bound $\sigma_m/m=1/(2\sqrt{N})$ has been derived, with $N$ the number of collected scattered photons. 


Our framework provides the means to quantitatively compare the theoretical performance of different microscopy techniques, with the possibility to take into account imaging artefacts, aberrations, or even other sources of noise. This allows researchers to undoubtedly decide whether or not a given technique is up for the task at hand. Furthermore, our results give the means to revisit the design of an experimental setup for optimized sensitivity. For example, the fundamental bounds derived here for axial localization are more precise than the ones obtained for fluorescence microscopy, showing the benefit of exploiting a coherent scattering process.

We finally note that this work has focused on the information contained in the measurements. The ability to precisely measure particle mass or localization also hinges on efficient estimators, which remains a challenging question to tackle \cite{iSCAT_tracking}. 
For this direction, one could benefit from the considerable amount of prior studies on localization for fluorescence microscopy~\cite{aguet2005maximum,mortensen2010optimized,sage2019super,nehme2020deepstorm3d}.

Another interesting direction for future studies is the consideration of more complex incidient light fields. While our results are derived assuming a plane incoming wave, exciting possibilities arise when considering adaptive wave-front~\cite{Juffmann2018, PhysRevApplied.15.024047, bouchet2021maximum} and amplitude~\cite{balzarotti_nanometer_2017} shaping of the input light, non-classical states~\cite{Liu2021, Pirandola2018}, or cavity enhanced measurement geometries~\cite{juffmann2016, nimmrichter2018, mader2015a, delhougne2021deeply}.

\medskip

\noindent{\textbf{Funding.}} 
This project has received funding from the European Research Council (ERC) under the European Union’s Horizon 2020 research and innovation programme (Grant Agreement No 758752). CB acknowledges funding from the FFG (project number 870337). JD acknowledges funding from European Research Council (ERC) under the European Union’s Horizon 2020 research and innovation programme (Grant Agreement No. 692726 GlobalBioIm). 

\medskip

\noindent{\textbf{Acknowledgments.}} The authors would like to thank Dorian Bouchet, Reza Gholami Mahmoodabadi, and Michael Unser for helpful discussions.

\medskip

\noindent{\textbf{Disclosures.}} The authors declare no conflicts of interest.

\medskip

\noindent The MATLAB code for this project can be found at \cite{code}. 

\appendix


\bibliography{refs}

\section{Vectorial PSF derivation}

In this section, we explicitly derive the PSF model used for all image simulations in this manuscript. This derivation is equivalent to the one found in~\cite{GholamiMahmoodabadi:20}, but we will discuss the different steps in more detail to obtain the 3D vectorial PSF formalism. 

\subsection{Design setting of the microscope}

\begin{figure}[t]
    \centering
    \includegraphics[width=\columnwidth]{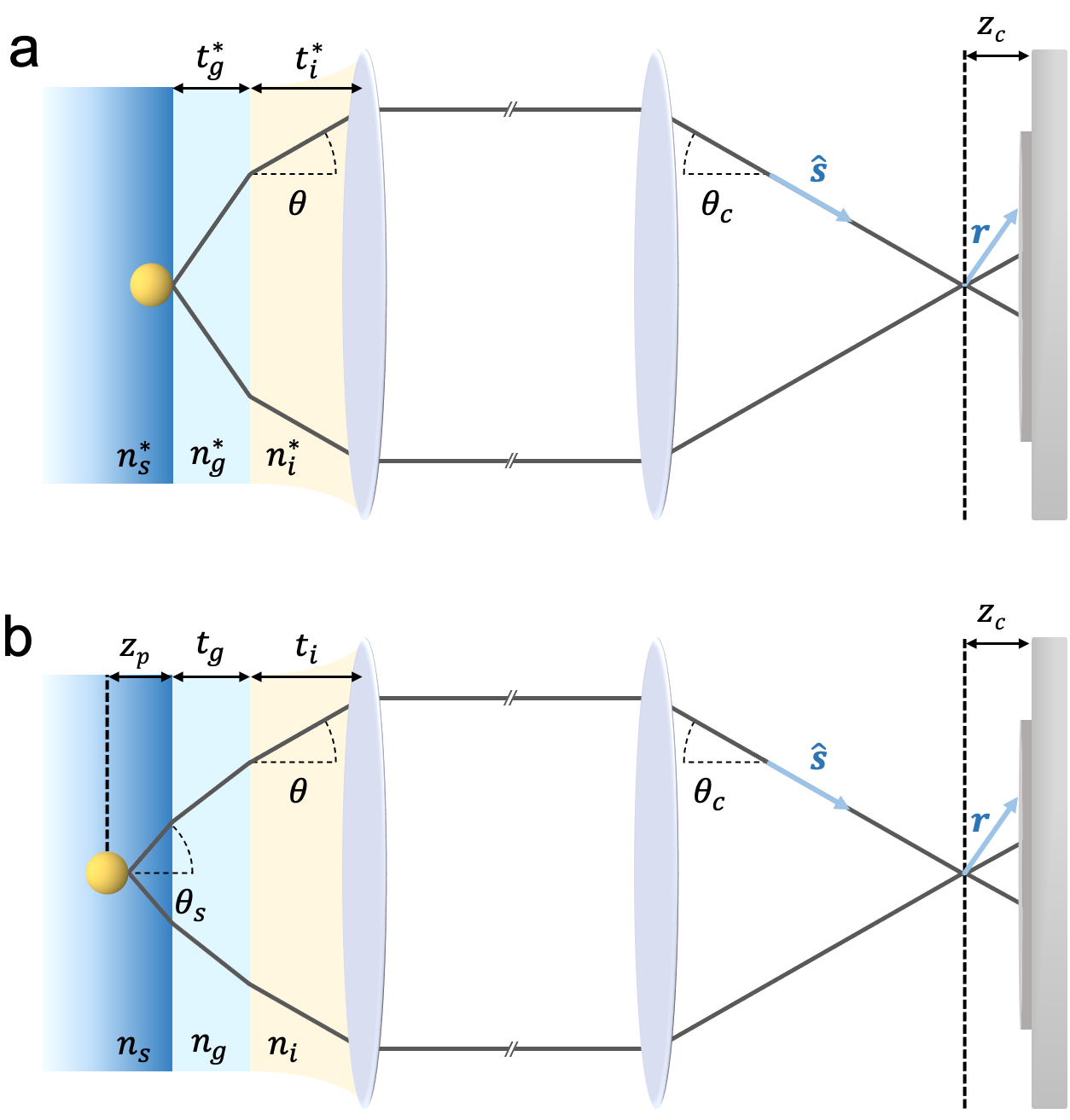}
    \caption[Description]{ Illustration of the layer structure for (a) the design parameters and (b) the abberrated case.}
    \label{fig: sketch_detail}
\end{figure}

We consider a three-layer model which describes the actual experiment~\cite{Gibson:92}, introducing three different regions with different refractive indices after the microscope objective: immersion oil, coverslip made of glass, and the sample medium. These three layers are described by their thickness and refractive index $(t_i, n_i)$, $(t_g, n_g)$, and $(z_p, n_s)$, respectively. 

Microscope objectives are engineered to produce aberration-free images of a specific plane. This is called the design setting of the objective, characterized by parameters $(t_i^*, n_i^*)$ and $(t_g^*, n_g^*)$. The imaged plane in this design setting is located right at the coverslip, which makes it independent of the sample parameters. Experiments are performed as close as possible to these design conditions. For example, we assume that the refractive index of the immersion oil and the parameters describing the cover glass match the design condition, i.e. $n_i = n_i^*$, and $(t_g, n_g) = (t_g^*, n_g^*)$. We also assume that $n_i=n_g$, such that reflections at the oil glass interface can be neglected. On the other hand, $(z_p, n_s)$ depend on the sample and the scatterer position, and the immersion oil thickness $t_i$ is a parameter set experimentally, related to the focus position of the microscope~\cite{Gibson:92}. 

It is in this setting that \cite{GholamiMahmoodabadi:20} have derived a vectorial PSF model following a similar approach for fluorescence microscopy  ~\cite{Aguet:09}. It is based on the vectorial model for a two-layer setting~\cite{Torok:95}, which showed excellent agreement with a model based on the Huygens-Fresnel principle~\cite{https://doi.org/10.1046/j.1365-2818.1999.00462.x}. The geometric aberrations of the three-layer setup can be desribed following~\cite{Gibson:92}, leading to a model that describes iSCAT experiments with high-fidelity~\cite{GholamiMahmoodabadi:20}.

We will start by describing the vectorial PSF model in the design setting, introducing the geometric aberrations of iSCAT and COBRI experiments in a second step. 

\subsection{The Richards-Wolf integral}

To model the 3D electric field propagation in this imaging configuration, we will use the Richards-Wolf integral approach~\cite{richards1959electromagnetic}. This vectorial model is well-suited for the high-NA configurations of localization techniques. From the electric strength vector at the back focal plane of the objective $\boldsymbol{\tilde{E}}_a$, the electric field in the object space can be calculated as:
\begin{equation}
    \boldsymbol{E}^b(\boldsymbol r) =
        - \frac{i k}{2 \pi} \int_0^{\alpha_{a}} \int_0^{2 \pi}
        \boldsymbol{\tilde{E}}^a
        e^{i k \hat{\boldsymbol{s}} \cdot \boldsymbol{r}}
        \sin \theta \sqrt{\cos \theta} d\theta d\phi
    \label{eq: app rw integral 0}
\end{equation}
where $\boldsymbol{r}$ is defined relative to the focus position and the Fourier space is parametrized by the polar angle $\theta$ and the azimuthal angle $\phi$. 

The electric field is decomposed into a superposition of plane waves $e^{i k \hat{\boldsymbol{s}} \cdot \boldsymbol{r}}$. The unit vector $\hat{\boldsymbol{s}}$ defining the direction of propagation has coordinates $(\sin \theta_c \cos \phi, \sin \theta_c \sin \phi, \cos \theta_c)$. The coordinates of the detector position are defined as $\boldsymbol{r} = (r \cos \phi_d, r \sin \phi_d, z_c)$. 
The equations above allow calculating the fields in different detector planes. In the following we will assume $z_c = 0$ for conciseness. The plane wave factor thus writes:
\begin{equation}
    e^{i k \hat{\boldsymbol{s}} \cdot \boldsymbol{r}} = e^{i k r \sin\theta_c \cos(\phi-\phi_d)} = e^{i k n_i r \sin\theta \cos(\phi-\phi_d)}
\end{equation}
where we have used Snell's law: $\sin \theta_c = n_i \sin \theta$.


\subsection{3D interferometric PSF in the design setting}

We start this discussion by considering a typical experiment in the design setting, where the nanometric scatterer is placed on the surface of the coverslip. There are two contributions to the measured electric field to consider: the reference wave and the scattered field from the nanoparticle generated by the dipole emission induced by the incident light. 

We assume the incident light $\boldsymbol{E}^0$ to be linearly polarized along the $x$ direction. In iSCAT, it is partially back-reflected at the glass-sample interface, yielding the reference field:
\begin{equation}
    \boldsymbol{E}^{\rm{ref}^*} = r_\parallel E^0 \hat{\boldsymbol e}_x
    \label{eq: app ref design}
\end{equation}
where $r_\parallel$ is the Fresnel reflection coefficient of the glass-sample interface for $p$-polarized waves at normal incidence:
\begin{equation}
    r_\parallel = \frac{n_g \cos \theta_s - n_s \cos \theta_g}{n_g \cos \theta_s + n_s \cos \theta_g}
\end{equation}
This factor may introduce a phase shift for the reflected light, but it does not in a typical iSCAT experiment, where $n_g$ and $n_s$ are real and $n_g>n_s$. For COBRI, the reference field is instead given by:
\begin{equation}
    \boldsymbol{E}^{\rm{ref}^*} = \beta t_\parallel E^0 \hat{\boldsymbol e}_x
    \label{eq: app ref design COBRI}
\end{equation}
with $t_\parallel$ the associated Fresnel transmission coefficient and $\beta$ an optional attenuation factor. 

The portion of light that is transmitted induces a dipole on the nanoparticle. The polarizability of a spherical particle of radius $a$, where $a \ll \lambda$, is given by:
\begin{equation}
    \alpha = 4 \pi a^3 \frac{\epsilon_1 - \epsilon_s}{\epsilon_1 + 2 \epsilon_s}
\end{equation}
where $\epsilon_1$ and $\epsilon_s$ are the permittivities of the nanoparticle and the sample medium respectively. The scattering phase and amplitude are given by:
\begin{equation}
    \psi_0 = \operatorname{arg}(\alpha)
\end{equation}
and:
\begin{equation}
    E^1 = \eta \frac{k^2}{\sqrt{6 \pi}} |\alpha| E^{\mathrm{inc}}
\end{equation}
where $\eta = 1/\pi \arcsin(\operatorname{min}(\text{NA}/n_s, 1))$ is a factor taking into account the light collection efficiency \cite{GholamiMahmoodabadi:20}. The incident field on the particle is $E^{\mathrm{inc}} = t_\parallel E_0$. Note that the near-fields of the radiating dipole are neglected here.

To propagate this dipole emission to the back focal plane, we consider the $p$ (denoted by $\parallel$) and $s$ (denoted by $\bot$) polarization components separately, determined by the following unit vectors:
\begin{align}
    \hat{\boldsymbol e}_\parallel^s &= (\cos \theta_s \cos \phi, \cos \theta_s \sin \phi, - \sin \theta_s) \\
    \hat{\boldsymbol e}_\bot &= (- \sin \phi, \cos \phi, 0)
\end{align}
The scattered field in the back focal plane then reads:
\begin{align}
    \tilde{\boldsymbol{E}}^{1^*}
        &= E^1 e^{i \psi_0} 
            \left((\hat{\boldsymbol e}_x \cdot \hat{\boldsymbol e}_\parallel^s) t_\parallel \hat{\boldsymbol e}_\parallel^s + 
            (\hat{\boldsymbol e}_x \cdot \hat{\boldsymbol e}_\bot^s) t_\bot \hat{\boldsymbol e}_\bot\right) \\
        &= E^1 e^{i \psi_0} \boldsymbol{A}
\end{align}
with:
\begin{equation}
    \boldsymbol{A} =
    \begin{pmatrix}
        t_\parallel \cos \theta_s \cos^2 \phi + t_\bot \sin^2 \phi \\
        (t_\parallel \cos \theta_s - t_\bot) \cos \phi \sin \phi
    \end{pmatrix}
\end{equation}

One can then use the Richards-Wolf integral to go from the back focal plane to the camera plane.
\begin{equation}
    \boldsymbol{E}^{\rm{cam}^*} =
    \boldsymbol{E}^{\rm{ref}^*} +
    \boldsymbol{E}^{\rm{sca}^*}
\end{equation}
with:
\begin{equation}
    \boldsymbol{E}^{\rm{sca}^*} =
        - \frac{i k}{2 \pi} E^1 e^{i \psi_0} \int_0^{\alpha_{a}} \int_0^{2 \pi} \boldsymbol{A}
        e^{i k \hat{\boldsymbol{s}} \cdot \boldsymbol{r}}
        \sin \theta \sqrt{\cos \theta} d\theta d\phi
\end{equation}
Note that since they were generated at the same plane, no phase retardation is introduced between the two waves. Also, for simplicity, the particle is placed on the optical axis. For a translated particle at position $\boldsymbol{r}_p = (x_p, y_p, 0)$, we have to replace $\boldsymbol{r}$ by $\boldsymbol{r} - \boldsymbol{r}_p$ 
in Eq.~\eqref{eq: app rw integral 0}. This adds a factor $e^{-ikn_i (x_p\cos\phi+y_p\sin\phi)\sin\theta}$ in the integrand of the previous equation that will be omitted for the rest of this derivation.

\subsection{3D aberrated iPSF}

Once the microscope is defocused or the particle is not directly bound to the cover slip, this leads to changes in optical path lengths that are described by an aberration function $\Lambda_s$, which depends on the coordinates in Fourier plane~\cite{Torok:95,haeberle2003focusing}. 
They can be expressed as \cite{Gibson:92}:
\begin{align}
    \Lambda_s(\theta) &= \frac{z_p n_s}{\cos \theta_s} + \frac{t_i n_i}{\cos \theta} - \frac{t_i^* n_i}{\cos \theta} \nonumber \\
    & \quad - n_i \sin \theta (z_p \tan \theta_s + t_i \tan \theta - t_i^* \tan \theta) \\
    \label{eq: gibson_lanni}
    &= z_p n_s \cos \theta_s + (t_i-t_i^*) n_i \cos \theta
\end{align}

The immersion oil thickness $t_i$ in the actual experiment is a parameter which is not easily measured experimentally. It can be estimated from the best geometric focus position $z_f$ thanks to the following relation \cite{Gibson:92}:
\begin{equation}
    t_i =  z_p - z_f + n_i \left(\frac{t_i^*}{n_i} - \frac{z_p}{n_s}\right)
\end{equation}

In the interferometric scheme of iSCAT, we also need to account for the additional phase due to the changed propagation distance of the incident wave:
\begin{equation}
    \Lambda_{s}' = n_s z_p + n_i (t_i - t_i^*)
\end{equation}
This term only corresponds to the way in, as for the backward pass it has already been accounted for in the aberration term. This term does not appear in COBRI, for which $\Lambda_{s}' = 0$ in the equations below. 

For the reference wave, the optical path difference compared to the design setting in iSCAT is given by:
\begin{equation}
    \Lambda_r' = 2 n_i (t_i - t_i^*)
\end{equation}
For COBRI, this difference is:
\begin{equation}
    \Lambda_r' = n_i (t_i - t_i^*) + n_s z_p
\end{equation}

We thus obtain the following equations: 
\begin{equation}
    \boldsymbol{E}^{\rm{cam}} =
    \boldsymbol{E}^{\rm{ref}} +
    \boldsymbol{E}^{\rm{sca}}
\end{equation}
with:
\begin{align}
    \boldsymbol{E}^{\rm{ref}} &= \boldsymbol{E}^{\rm{ref}^* }
    \label{eq: app ref phase}\\
    \boldsymbol{E}^{\rm{sca}} &= - \frac{i k}{2 \pi} E^1 e^{i \psi_0} \times
    \nonumber\\
    & \qquad 
    \int_0^{\alpha_{a}} \int_0^{2 \pi} \boldsymbol{A}
    e^{i k \Lambda}
    e^{i k \hat{\boldsymbol{s}} \cdot \boldsymbol{r}}
    \sin \theta \sqrt{\cos \theta} d\theta d\phi
    \label{eq: app rw integral}
\end{align}
where we have concatenated all the aberration terms into $\Lambda = \Lambda_s + \Lambda_s' - \Lambda_r'$. 
Since the intensities only depend on the phase difference between scattered and reference waves, and since $\Lambda_r'$ does not depend on $\theta$, the optical path length $\Lambda_r'$ term of the reference has been introduced in the Richards-Wolf integral. 
We obtain the expressions given in the main text, namely:
\begin{equation}
    \Lambda = n_s z_p (\cos \theta_s+1) + n_i (t_i-t_i^*) (\cos \theta - 1)
\end{equation}
for iSCAT, and this term for COBRI is:
\begin{equation}
    \Lambda = z_p n_s (\cos \theta_s-1) + n_i (t_i-t_i^*) (\cos \theta-1)
\end{equation}

\subsection{Compact formalism}

An integration of Eq.~\eqref{eq: app rw integral} over $\phi$ leads to the final expression:
\begin{equation}
    \boldsymbol{E}^{\rm{sca}} = - \frac{i k E^1 e^{i \psi_0}}{2}
    \begin{pmatrix}
        I_0+I_2\cos(2\phi_d) \\
        I_2 \sin(2\phi_d)
    \end{pmatrix}
\end{equation}
where:
\begin{align}
    I_0 &= 
    \int_0^{\alpha_{a}} B_0(\theta) \left(t_\bot + 
        t_\parallel \frac{1}{n_s} \sqrt{n_s^2-n_i^2 \sin^2\theta} \right) d\theta
    \\
    I_2 &= 
    \int_0^{\alpha_{a}} B_2(\theta) \left(t_\bot - 
        t_\parallel \frac{1}{n_s} \sqrt{n_s^2-n_i^2 \sin^2\theta} \right) d\theta
\end{align}
with:
\begin{equation}
    B_m(\theta) = \sqrt{\cos \theta} \sin \theta J_m(n_i k r \sin \theta) e^{ik \Lambda}
\end{equation}
using $J_m$ the Bessel function of order $m$ for $m = 0, 2$. 
To compute this integration, Snell's law has been used, $\cos \theta_s = 1 / n_s \times \sqrt{n_s^2 - n_i^2 \sin^2 \theta}$, along with the following identities \cite{richards1959electromagnetic}:
\begin{align}
    \int_0^{2 \pi} \cos(m \phi) e^{i \rho \cos(\phi - \phi_d)} d\phi &= 
        2 \pi i^m J_m(\rho) \cos(m\phi_d) \\
    \int_0^{2 \pi} \sin(m \phi) e^{i \rho \cos(\phi - \phi_d)} d\phi &= 
        2 \pi i^m J_m(\rho) \sin(m\phi_d)
\end{align}
This compact formalism only requires the computation of a single integral, compared to the two integrations in Eq.~\eqref{eq: app rw integral}. This accelerates the numerical computation of the PSF for all camera pixel positions $\boldsymbol{r}$. 

\subsection{Normalization of the photon number}

The Fisher Information is proportional to the number of photons $N$, while the CRB scales as $1/N$ and the minimal standard deviation as $1/\sqrt{N}$. This photon number is proportional to:
\begin{equation}
    N \propto \iint d\boldsymbol{r} |\boldsymbol{E}^{\rm{sca}}|^2
\end{equation}
where we can neglect a constant prefactor. 

The Cram\'er-Rao bounds presented in this work are normalized for one photon going through the imaging system. We thus normalize all electric fields with a global prefactor such that $N = 1$, similar to \cite{backlund2018fundamental}. 

\section{Fundamental bounds using quantum CRB}


In this section we derive fundamental bounds on the localization precision of the nanoparticle that can be achieved in interferometric imaging schemes. This work is related to \cite{backlund2018fundamental} where fundamental bounds have been derived for fluorescence microscopy and \cite{bouchet2021maximum} where these quantities have been computed for coherent states.



\subsection{Fundamental bound on localization}

We consider a nanoparticle at position $(x_p, y_p, z_p)$. For brevity, the Richards-Wolf integral model described in Eq.~\eqref{eq: app rw integral} can also be written as:
\begin{equation}
    \boldsymbol{E}^{\rm{sca}} = 
    \int_0^{\alpha_{a}} \int_0^{2 \pi} \boldsymbol{\epsilon}(\theta, \phi) 
    e^{i k \hat{\boldsymbol{s}} \cdot \boldsymbol{r}} d\theta d\phi
    \label{eq: app abberated RW integral for quantum}
\end{equation}
with the amplitude given by:
\begin{align}
    \boldsymbol{\epsilon}(\theta, \phi) &= - \frac{ik}{2\pi} E^1 e^{i \psi_0} \sin\theta \sqrt{\cos\theta} \times \nonumber \\
    &\qquad e^{ik\Lambda} e^{-ikn_i (x_p\cos\phi+y_p\sin\phi)\sin\theta} \boldsymbol{A}
\end{align}
The scattered field in the back focal plane of the objective is described as a superposition of coherent states in a basis of plane waves parametrized by $(\theta, \phi)$, and with an amplitude $\epsilon(\theta,\phi)$. 
We normalize this amplitude to one scattered photon passing through the imaging system: 
\begin{equation}
\int_0^{\alpha_{a}} d\theta \int_0^{2\pi} d\phi |\boldsymbol{\epsilon}(\theta, \phi)|^2 = 1
\end{equation}

To emphasize the dependence on the particle position, we note that the amplitude of the coherent state is of the following form:
\begin{equation}
    \boldsymbol{\epsilon}(\theta, \phi) = 
        \boldsymbol{E}^s(\theta, \phi) e^{i k n_i ((x_p \cos \phi + y_p \sin \phi) \sin \theta) + ik n_s z_p (\cos \theta_s + \xi)}
\end{equation}
with $\boldsymbol{E}^s(\theta, \phi)$ independent from $(x_p, y_p, z_p)$. In this expression, we have neglected the reference and the vectorial formalism, which would add an additional dimension. The $\xi$ term in the $z_p$ term is linked with the geometrical path length difference caused by the particle height $z_p$. It is equal to +1 for iSCAT and -1 for COBRI.

The QFI of a pure state made of orthogonal coherent states has a very concise formulation, as derived in Eq. (1) of \cite{bouchet2021maximum}. Thus, we obtain the Quantum Fisher Information for each parameter:
\begin{align}
    \label{eq: app qcrb x}
    \mathcal{K}_{xx} &= 
        4 \int_0^{\alpha_{a}} d\theta \int_0^{2\pi} d\phi
        \left|\partial_x \boldsymbol{\epsilon}(\theta, \phi)
        \right|^2\\
    &= 4 \int_0^{\alpha_{a}} d\theta \int_0^{2\pi} d\phi
        \left|\boldsymbol{\epsilon}(\theta, \phi) k n_i \cos \phi \sin \theta
        \right|^2
\end{align}
Similarly:
\begin{align}
    \mathcal{K}_{yy} &= 4 \int_0^{\alpha_{a}} d\theta \int_0^{2\pi} d\phi
        \left|\boldsymbol{\epsilon}(\theta, \phi) k n_i \sin \phi \sin \theta
        \right|^2 \\
    \mathcal{K}_{zz} &= 4 \int_0^{\alpha_{a}} d\theta \int_0^{2\pi} d\phi
        \left|\boldsymbol{\epsilon}(\theta, \phi) k n_s (\cos \theta_s \pm 1)
        \right|^2
\end{align}
where, following from Snell's law, $\cos\theta_s = \sqrt{1-n_s^2\sin^2\theta/n_i^2}$. Injecting the expression of $\boldsymbol{\epsilon}$ yields the final formula of the QCRB described in Eqs.~\eqref{eq: qcrb x} to \eqref{eq: qcrb z}.

Finally, the associated Quantum Cram\'er Rao bounds are:
\begin{align}
    \sigma_{\mathrm{QCRB},\,x}^2 &= 
        \frac{1}{\mathcal{K}_{xx}} \\
    \sigma_{\mathrm{QCRB},\,y}^2 &= 
        \frac{1}{\mathcal{K}_{yy}} \\
    \sigma_{\mathrm{QCRB},\,z}^2 &= 
        \frac{1}{\mathcal{K}_{zz}}
\end{align}
The integrals on $\theta$ and $\phi$ are computed numerically and the resulting QCRBs are shown in Fig.~\ref{fig: localization_precision}.

\subsection{Fundamental bound on mass photometry}

The amplitude of the coherent states is proportional to polarizability, and therefore to mass $m$. We write $\epsilon(\theta, \phi) = m \delta(\theta, \phi)$, where the introduced quantity $\delta(\theta, \phi)$ does not depend on $m$. 

The QFI is thus given by: 
\begin{align}
    \mathcal{K}_{mm} &= 
        4 \int_0^{\alpha_{a}} d\theta \int_0^{2\pi} d\phi
        \left|\partial_m \boldsymbol\epsilon(\theta, \phi)
        \right|^2\\
    &= 
        4 \int_0^{\alpha_{a}} d\theta \int_0^{2\pi} d\phi
        \left|\boldsymbol\delta(\theta, \phi)\right|^2 \\
    &= \frac{4}{m^2}
\end{align}
thanks to the normalization of $\boldsymbol{\epsilon}$.

The fundamental bound on mass estimation then writes:
\begin{equation}
    \sigma_{\mathrm{QCRB},\,m}^2 = 
        \frac{1}{\mathcal{K}_{mm}} = \frac{m^2}{4}
\end{equation}

\section{Parameters used for simulations}

For all simulations the following parameters were used: We considered a wavelength of 517.5 nm and used a numerical aperture of 1.3. The refractive indexes were: $n_s = 1.33$, $n_g = 1.5$ and $n_i = 1.5$. We took the ideal scenario where $n_i = n_i^*$ and $t_g = t_g^*$. The permittivity of the gold nanoparticle was set to $-3.7328+ 2.7725i$  \cite{PhysRevB.6.4370}. The thicknesses used were, $t_i^* = 100$~{\textmu}m, $t_g^* = 170$~{\textmu}m and $t_i$ was given by Equation (A17). The particle position was centered at $x_p = y_p = 0$ for all simulations. Gold nanoparticles had a 30 nm diameter with a density of 19.3 $\times$ $10^{3}$~kg/m$^3$. 
\end{document}